\documentstyle[12pt]{article}
\begin{document}
\title{Wilson Loop and the Treatment of Axial Gauge Poles}
\author{{Satish. D. Joglekar}\thanks{e-mail:sdj@iitk.ac.in}\\
Department of Physics, Indian Institute of Technology\\
 Kanpur 208 016, UP, India\\
{A. Misra} \thanks{e-mail:aalok@iitk.ac.in}\\
Institute of Physics, Sachivalaya Marg\\
Bhubaneswar 751 005, Orissa, India}
\maketitle
\begin{abstract}
We consider the question of gauge invariance of the Wilson loop in the light of a 
new treatment of axial gauge propagator
proposed recently based on a finite field-dependent
BRS (FFBRS) transformation. We remark
that as under the FFBRS transformation the vacuum expectation
value of a gauge invariant observable remains unchanged,
our prescription automatically satisfies the Wilson loop criterion.
Further, we give an argument for {\it direct} verification
of the invariance of Wilson loop to O$(g^4)$ using the earlier work
by Cheng and Tsai. We also note that
our prescription preserves the thermal Wilson loop to O$(g^2)$.
\end{abstract}

The axial-type gauges have found favor in many Standard Model calculations on account
of their formal ghost-free nature. Associated, however, with the axial
gauges, are the problems posed by the spurious 
$1/(\eta\cdot q)^p$-type singularities. Much literature
has been devoted to the question of how these
these singularities can be interpreted \cite{diff}. There are a number
of prescriptions for these singularities, some of which
are the principle-value prescription (PVP) \cite{pvp}, Leibbrandt-Mandelstam
(LM)\cite{leibmand}, the $\alpha$-prescription
\cite{landsh}, the van Niewenhuizen-Landshoff prescription \cite{landshniew}
(for a subset of axial gauges), etc.
PVP and LM are associated with extensive
literature and some of their limitations have been established \cite{ref1}.
Many of the prescriptions have been of an ad-hoc nature and as a result
require validation in the form of various checks such as validity of 
WT identities \cite{diff}, preservation of Wilson
loop \cite{chengtsai} and the thermal Wilson loop \cite{nadk}, etc.
It has, for example, been shown that
the PVP does not preserve the Wilson loop to O$(g^4)$\cite{ccm}.
The same holds for the thermal Wilson loop and the $\alpha$-prescription
\cite{lm}.

Recently, we have proposed
\cite{BRSpap1,BRS2lett}, a direct way to handle
the $1/(\eta\cdot q)^p$-type singularities based on a 
``finite field-dependent BRS (FFBRS) transformation"\cite{jm,jb}.
This method is based directly on correlating the generating functionals themselves
\cite{BRS3} in the two set of gauges via FFBRS transformation. Such a set of 
transformations automatically preserves the vacuum expectation value of gauge-invariant
observables to all orders. To be concrete, consider the effective actions
(in obvious notations)
\begin{eqnarray}
\label{eq:SeffL}
& & S_{\rm eff}^L[A,c,\bar c,\epsilon]=S^0[A]-\int d^4x\biggl({1\over{2\lambda}}(\partial\cdot A^\alpha)^2
+{\bar c}^\alpha M^{\alpha\beta}c^\beta+i\epsilon[{1\over2}(A^\alpha)^2-{\bar c}c]\biggr)
\nonumber\\
& & \equiv S^L_{\rm eff}+\epsilon O_1[A,c,{\bar c}],
\end{eqnarray}
where the $\epsilon$-terms in $S_{\rm eff}$ have been 
put in accordance with the replacement
$q^2\rightarrow q^2+i\epsilon$ for all poles in the Lorentz-type gauges
\cite{gthooft}. As implied in \cite{BRS2lett,BRS3}, the correct treatment 
of axial-type gauges
is obtained via an effective action (in obvious notations)
\begin{equation}
\label{eq:SeffA}
S_{eff}^A[A^\prime,c^\prime,{\bar c}^\prime,\epsilon]=
S^0[A^\prime]-\int d^4x\biggl({1\over{2\lambda}}(\eta\cdot A^{\prime\alpha})^2+
{\bar c}^{\prime\alpha} \tilde M^{\alpha\beta}c^{\prime\beta}\biggr)
+\epsilon O^\prime_1[A^\prime,c^\prime,{\bar c}^\prime]
\end{equation}
where the $O(\epsilon)$ terms in $S^A_{\rm eff}$ are obtained by
performing a field transformation
$[A,c,{\bar c}]\rightarrow [A^\prime,c^\prime,{\bar c}^\prime]$
which is an FFBRS transformation
\cite{BRS2lett,jm} such that it preserves the vacuum-to-vacuum
amplitude:
\begin{equation}
\label{eq:vactovac}
\int {\cal D}A{\cal D} c{\cal D}{\bar c}e^{iS^L_{\rm eff}[A,c,{\bar c}]}
=\int {\cal D}A^\prime {\cal D}c^\prime{\cal D}{\bar c}^\prime
e^{iS^A_{\rm eff}[A^\prime, c^\prime,{\bar c}^\prime]}.
\end{equation}
Such field transformations were constructed
in a general context in \cite{jm}. As remarked in \cite{BRSpap1,BRS3}, these
transformations also  preserve the vacuum expectation values of gauges-invariant
observables:
\begin{equation}
\label{eq:GIvev}
\int {\cal D}A{\cal D} c{\cal D}{\bar c}O^{GI}[A]e^{iS^L_{\rm eff}[A,c,{\bar c}]}
=\int {\cal D}A^\prime {\cal D}c^\prime{\cal D}{\bar c}^\prime
O^{GI}[A^\prime]e^{iS^A_{\rm eff}[A^\prime, c^\prime,{\bar c}^\prime]}.
\end{equation}
Such a procedure was indeed adopted for the purpose of interpretation
of ${1\over{(\eta\cdot q)^p}}$-type singularities in
axial-type gauges \cite{BRSpap1,BRS2lett}.

In view of the fact that the Wilson loop
\begin{equation}
\label{eq:W[L]}
W[L]=\langle {\rm Tr} P T e^{-ig\oint A^\mu dx_u}\rangle
\end{equation}
is a gauge-invariant observable, it follows from
(\ref{eq:GIvev}) that our treatment is such that by its very construction,
the Wilson loop $W[L]$ has the same value in the Lorentz and axial-type 
gauges to {\it all} orders.
 
While (\ref{eq:GIvev}) provides a formal proof of the equivalence of
$W[L]$ for the two sets of gauges, we shall also
provided a reasonably brief proof of this statement to
O$(g^4)$. We utilize the work of Cheng and Tsai \cite{chengtsai} in order
to shorten the argument.

We note that our propagator of \cite{BRSpap1, BRS2lett}
is in fact of the general form used by Cheng
and Tsai \cite{chengtsai}:
\begin{equation}
\label{eq:prop}
\tilde D_{\mu\nu}={-i\over{k^2+i\epsilon^\prime}}(g_{\mu\nu}-a_\mu k_\nu
-b_\nu k_\mu),
\end{equation}
where $a_\mu(k)=-b_\mu(-k)$ by Bose symmetry. Cheng and Tsai \cite{chengtsai}
prove the equivalence of Wilson loops in two gauges by introducing an effective ghost-ghost-gluon
vertex:
\begin{equation}
\label{eq:ghghglver}
G^\mu(k)=i{[(a\cdot k)-1]k^\mu-k^2a^\mu(k)]\over{k^2+i\epsilon^\prime}}
\end{equation}
(which includes propagator of outgoing ghost). As such
the entire proof of equivalence of the Wilson loops
in the Lorentz and axial-type gauges would go through,
were it not for the fact that in our treatment the ghost as well as
gluon propagator arise rather out of the effective
action $S^A_{\rm eff}[A^\prime,c^\prime,{\bar c}^\prime]$.
Consequently, the results of \cite{chengtsai}
would imply the equivalence of Wilson loop
in the two sets of gauges (axial and Lorentz) 
to O$(g^4)$ verified there if we could show that
the two treatments of ghosts do not make any difference to this order.

For the sake of brevity, we refer the reader 
to the diagrams (Fig. 1) in \cite{chengtsai}.
We shall prove the desired result simply by  showing   that the
(only) one diagram required to be considered in this
calculation in fact vanishes for all loops for $\eta^2<0$ and in fact 
vanishes for a large class of loops for $\eta^2>0$.

The only diagram to O$(g^4)$ that is relevant is the diagram 
1(b) of \cite{chengtsai} which is a diagram with two ends of a gluon
propagator with a ghost-loop self-energy insertion. The entire propagator
diagram (with the two gluon propagators) has the structure:
\begin{equation}
\label{eq:tenstruc}
G_{\mu\nu}(p,\eta)=Ap_\mu p_\nu+Bp_\mu\eta_\nu +Cp_\nu\eta_\mu+D\eta_\mu\eta_\nu
\end{equation}
where $A,B,C$ and $D$ are functions of $p^2,\eta\cdot p$ 
(and also depend on $\eta^2$). The whole contribution of this diagram
to the Wilson loop is then of the form:
\begin{equation}
\label{eq:loopcontrib}
W^{[1]}\equiv\int d^4p\oint dx^\mu\oint dy^\nu e^{ip\cdot(x-y)}G_{\mu\nu}(p,\eta).
\end{equation}  
We shall now give
a simple argument that for all loops the first three terms in 
(\ref{eq:tenstruc}) do not contribute to (\ref{eq:loopcontrib}). 
We note that the contribution of the first two terms to 
(\ref{eq:loopcontrib}) is of the form
\begin{equation}
\label{eq:ABtype1}
W^{[1]}=\int d^4p\oint d x^\mu\oint dy^\nu e^{ip\cdot(x-y)} p_\mu \tilde A_\nu(p^2,\eta\cdot p)
\end{equation}
which could be re-expressed as:
\begin{eqnarray}
\label{eq:ABtype2}
& & W^{[1]}=\int d^4p\tilde A_\nu(p^2,\eta\cdot p)\oint dy^\nu e^{-ip\cdot y}
\oint dx^\mu p_\mu e^{ip\cdot x}\nonumber\\
& & =\int d^4p \tilde A(p^2,\eta\cdot p)\oint d y^\nu e^{-ip\cdot y}
\oint d x^\mu (-i\partial_\mu)e^{ip\cdot x}.
\end{eqnarray}
Now, for a fixed $p$ and $y$, we can  evaluate
\begin{equation}
\label{eq:totderiv}
\oint dx^\mu\partial_\mu (e^{ip\cdot x})
\end{equation}
from `$y$' to `$y$' along the loop and find that it
vanishes. A similar argument applies to the third term:  Here we evaluate
$\oint dy^\nu p_\nu e^{-ip\cdot y}$ first.

We now consider the contribution of the last term. It reads
\begin{equation}
\label{eq:Dtype}
W^{[1,4]}\equiv\int d^4p\oint_L dx^\mu\eta_\mu e^{ip\cdot x}
\oint_L dy^\nu\eta_\nu e^{-ip\cdot y}D(p^2,\eta\cdot p,\eta^2).
\end{equation}
The Wilson loop is a gauge-invariant quantity. We expect that 
$W[L]$ is independent of the orientation of $\eta$.
This is because an infinitesimal change
in $\eta$ obtained through a Lorentz transformation, is in fact
an infinitesimal change in  the gauge fixing
term (and the associated change in the ghost term). Under such a change,
the use of BRS WT identities will imply  invariance of $W[L]$, the expectation
vale of a gauge invariant operator\cite{iz}.
Therefore, in evaluating $W[L]$, we can
choose to perform a suitable Lorentz transformation, 
$\eta^\mu\to\eta^{\prime\mu}=\Lambda^\mu_\nu\eta^\nu$ on $\eta^\mu$. Then
(\ref{eq:Dtype}) becomes:
\begin{equation}
\label{eq:Weta'eta}
W^{[1,4]}
=\int d^4p\oint_L dx^\mu\eta_\mu^\prime e^{ip\cdot x}
\oint_L dy^\nu\eta_\nu^\prime e^{-ip\cdot y} D(p^2,\eta^\prime\cdot p,\eta^2).
\end{equation}
Now, (\ref{eq:Weta'eta}) would
vanish for a planar loop if we could choose $\eta^{\prime\mu}=\Lambda^\mu_\nu\eta^\nu$
such that it is perpendicular to the loop L
($\eta^{\prime\mu}dx_\mu=0$). We now investigate when this is possible
for a planar loop L. The plane of the loop is described (not necessarily
uniquely) by the intersection of 
\begin{equation}
\label{eq:eqnsplanL}
a^\mu x_\mu=A\ {\rm and}\ b^\mu x_\mu=B.
\nonumber
\end{equation}
(i) Let $\eta^2<0$. We note that  at least one of $a^\mu, b^\mu$   and
$c^\mu\equiv (a^\mu-{a^0\over b^0}b^\mu)$ (with $b^0\neq 0$)
is necessarily spacelike. We pick a space-like  vector
of these three and call it 
$n^\mu$. Then for $\eta^2<0$, we can always choose $\eta^\prime$ parallel
to $n$. Evidently, since $d(n\cdot x)=0$, we have $\eta^{\prime\mu}d x_\mu=0$
on $L$ and (\ref{eq:Weta'eta}) vanishes.

(ii) Let $\eta^2>0$. Suppose it is possible to construct a time-like 
vector out of linear combinations of $a$ and $b$. Then, we call this
`$n$' and orient $\eta^\prime$ parallel to $n$. The argument proceeds
as above. Evidently, the argument in this case, does not hold for those planar loops
for which time-like normal does not exist.

(iii) Let $\eta^2=0$. If $(a\cdot b)^2>a^2b^2$, a real $\eta$ exists
such that $a+\beta b$ is light-like. In such case, we choose
$n=a+\beta b$ and the argument proceeds as before.

The above argument in fact applies to loops more general than a planar loop.
For example, for $\eta^2 < 0$, the argument applies to any loop confined to a
3-Dimensional subspace normal to some space-like vector.

The argument easily   extends itself to any arbitrary loop for 
$\eta^2<0$ and to a subclass of arbitrary loops for $\eta^2\geq0$. 
We close the loop
by a surface $S$ which we cover by a patchwork of N planar loops (We might require
the limit $N\to\infty$ for a given loop). We can then write
\begin{equation}
\label{eq:LsumCi}
W[L]=\langle {\rm Tr} T P {\rm exp}\biggl(\oint_L A_\mu dx^\mu\biggr)\rangle
=\langle {\rm tr}TP {\rm exp}\biggl(\sum_{i=1}^N\oint_{C_i}A_\mu dx^\mu\biggr)\rangle.
\end{equation}

The relevant O$(g^4)$ contribution to $W[L]$ from ghost diagrams comes from the
quadratic terms in the expansion of the exponent, for which the path
ordering is immaterial since ${\rm tr}(T^aT^b)={\rm tr}(T^bt^a)$.  Such a typical term
(apart from an overall coefficient) reads
\begin{equation}
\oint_{C_i}dx^\mu\oint_{C_j}dy^\nu G_{\mu\nu}(x,y).
\end{equation}
We can now apply the arguments following equation (\ref{eq:tenstruc})
to the above typical term. While  doing so, we shall come across contribution
analogous to (\ref{eq:Weta'eta}). In such a term, we need to choose 
$\eta^\prime$ perpendicular  to {\it one} of the two loops $C_i$ and
$C_j$. Then the entire argument goes through for each such term $(i,j)$
separately, provided $\eta^\prime$ can always be chosen normal to one
of the loops\footnote{The above argument in fact assumes that our argument applies
separately to each such term ($i,j$). This can be seen as follows: Let an arbitrary
closed loop $L$ be broken up as a sum of two arbitrary closed loops $L_1$ and
$L_2$. Then $W[L]$ can be broken up as $W[L]=W[L_1] + W[L_2]+W[L_1,L_2]$. Here,
$W[L_1,L_2]$ depends on both the loops. Now the entire argument of Cheng and Tsai and
the argument we gave for the independence of direction of $\eta$
applies to each
of $W[L], W[L_1]$ and $W[L_2]$ and therefore to $W[L_1,L_2]$. This argument can easily
be extended to the case when $L$ is broken up in to $N$ loops. Hence etc.}. 
As shown earlier, this is   always possible for $\eta^2<0$. 
For $\eta^2\geq0$,
this will hold only for a subclass of loops.

We also note that the O$(g^2)$ thermal Wilson loop considered in 
\cite{nadk} is  given by:
\begin{equation}
\label{eq:Wilthermal}
W_R=1+{(N_c^2-1)g^2\over{2N_cT}}\int {d^3k\over{(2\pi)^3}}
[1-cos(\vec k\cdot\vec R)]D_{00}(k^0=0,\vec k)
\end{equation}
and thus depends only on $D_{00}(k^0=0,\vec k)$. We find
that for the propagator in \cite{BRSpap1,BRS2lett},
$D_{00}(k^0=0,\vec k)={g_{00}\over{k^2+i\epsilon}}$ which
is the same as $D_{00}(k^0=0,\vec k)$  for Lorentz
gauges and as such $W_R$ has the same value as in Lorentz-type gauges.

To summarize, we have emphasized that the treatment we have given for axial 
gauge poles \cite{BRSpap1,BRS2lett} is by its very construction, compatible with the
preservation of the Wilson loop to all orders. We have further given the proof of this
statement to O$(g^4)$  using the earlier work by
Cheng and Tsai. Our proof holds for any arbitrary  loop for
$\eta^2<0$ and for a subclass of loops for $\eta^2\geq0$. We have
 also  noted the preservation of the thermal Wilson loop to O$(g^2)$
for our treatment.

\section*{Acknowledgement}

The authors thank Dr.M.Lavelle for a communication.


\begin{thebibliography}{99}
\bibitem{diff} See for example references in G.
Leibbrandt, Rev Mod Phys ${\bf 59}$, 1067 (1987);
G. Leibbrandt, {\it Noncovariant Gauges},
World Scientific (Singapore) 1994.
A.Bassetto, G.Nardelli and R.Soldatti,
{\it Yang Mills Theories in Algebraic Non-Covariant Gauges},
World Scientific (Singapore) 1991.
\bibitem{pvp} W. Kummer, Nucl Phys. B ${\bf 100}$, 106 (1976).
\bibitem{leibmand} Leibbrandt G, Phys Rev D ${\bf 29}$,
1699 (1984),\\
Mandelstam S, Nucl Phys B ${\bf 213}$, 149 (1983).
\bibitem{landsh} P.V.Landshoff Phys.Lett.B 169,69 (1986).
\bibitem{landshniew} P.V. Landshoff and P. von
Niewenhuizen, Phys. Rev. D ${\bf 50}$
4157 (1994).
\bibitem{ref1} See e.g.ref.1.
\bibitem{chengtsai} H.Cheng and Tsai,Phys.Rev.D36,3196(1987).
\bibitem{nadk} S.Nadkarni Phys. Rev. D33,3738(1986).
\bibitem{ccm} S.Caracciolo,G.Curci,P.Minotti Phys.Lett.B113,311(1982).
\bibitem{lm} M.Lavelle and D.McMullan Z.Phys. C 59,351(1993).
\bibitem{BRSpap1} Satish D. Joglekar and A. Misra, hep-th/9812101,to
appear in Jour.Math.Phys.(2000).
\bibitem{BRS2lett} Satish D. Joglekar and A. Misra,
Mod. Phys. Lett A {\bf 14}, No. 30, 2083  (1999).
\bibitem{jm} S.D.Joglekar and B.P.Mandal, Phys Rev D
${\bf 51}$, 1919 (1995).
\bibitem{jb} R.S.Bandhu and S.D.Joglekar
J.Phys.A-Math and General ${\bf 31}$, 4217 (1998);
S.D.Joglekar in {\it Finite Field-dependent
BRS (FFBRS) Transformations and Axial Gauges}; invited talk at the
conference titled {\it Theoretical Physics Today: Trends and
Perspectives} held at IAS Calcutta, April 1998;   Ind. J.
Phys.  {\bf 73B}(2), 137 (1999).
\bibitem{BRS3} Satish D. Joglekar and A. Misra,hep-th/9909123,
to appear in Int.J.Mod.Phys.A (1999).
\bibitem{gthooft}G.'t Hooft Nucl.Phys.B {\bf 33}, 173 (1971).
\bibitem{iz} {\it Qutantum Field Theory}, C.Itzykon, J-B.Zuber, McGraw Hill (1980).
\end{thebibliography}
\end{document}